\documentclass[11pt]{article}
\usepackage{amsmath,amssymb,amsfonts}
\usepackage{graphicx}
\usepackage{caption}
\usepackage[font={scriptsize}]{subcaption}
\newcommand{\RNum}[1]{\uppercase\expandafter{\romannumeral #1\relax}}
\newcommand{\be}{\begin{equation}}
\newcommand{\ee}{\end{equation}}
\newcommand{\bea}{\begin{eqnarray}}
\newcommand{\eea}{\end{eqnarray}}
\newcommand{\ba}{\begin{array}}
\newcommand{\ea}{\end{array}}
\newcommand{\M}{\mathcal{M}}
\newcommand{\N}{\mathcal{N}}
\newcommand{\D}{\mathcal{D}}
\newcommand{\K}{\mathcal{K}}

\newcommand{\htwo}{h_{2,1}}

\usepackage{hyperref}
\hypersetup{
   colorlinks   = true, 
    urlcolor     = blue, 
    linkcolor    = blue, 
    citecolor   = blue
}
\begin{document}

\begin{titlepage}

\begin{center}

\vspace{200pt}
%\bigskip

{\Large \bf The Cosmological Constant Problem and the Extra Dimensions of $\N=2$ $\D=5$ Supergravity} 

\bigskip

\bigskip

{\bf  Safinaz Salem}\\
\smallskip

{ \small \it  
 Department of Physics, Faculty of Science, Al Azhar University, Cairo 11765, Egypt\\

and

Center for Fundamental Physics, Zewail City of Science
and Technology, 6th of October City, Giza 12578, Egypt}

\bigskip

{\tt  safinaz.salem@azhar.edu.eg}

\bigskip

\vspace*{.5cm}

{\bf Abstract}\\
\end{center}
\noindent
We propose an interpretation for the cosmological constant problem based on modeling the universe as a 3-brane embedded in the bulk of 5-dimensional supergravity with hypermultiplets. When  solving the modified Friedmann equations the complex structure moduli of the Calabi-Yau manifold cancel the large value of the vacuum energy density on the brane, wherein this value transfers to an Anti-de Sitter fifth dimensional bulk compactified on a scale $ (\sim 10^{-26} ~\text{m}) $. An effective dark energy density is produced on the brane which equals the observed value responsible for the late-time acceleration of the universe. The bulk undergoes a decelerated contraction while we show that cosmic expansion of the brane-universe consists with the recent observed data of  the $\Lambda$CDM model. On the other hand, through this model, the hierarchy between the electroweak scale $(M_{EW}= 100 ~
\text{GeV})$ and the Planck scale $(M_{pl} \sim 10^{18} ~
\text{GeV})$ is interpreted by compactifying the 6-extra dimensions of $\D=5$ supergravity on a sub-millimeters scale, where the fundamental scale of gravity within these extra dimensions is $\M_{EW}$.

\bigskip

\bigskip

\end{titlepage}

\section{Introduction}
The paradox of the accelerated late-time expansion of the universe is one of the most controversial problems in modern cosmology. The cosmological constant in Einstein's field equations is related to the vacuum energy density and receives many quantum corrections as first suggested in 1967 \cite{Zel'dovich:1967, Rugh:2002}. This value magnificently exceeds the value of the dark energy density which is responsible for the universe's current accelerated expansion as observed in 1998 using distant type Ia supernovae \cite{Riess:1999} and the anisotropy of the cosmic microwave background (CMB) \cite{Planck:2018}.

%\vspace{1cm}

One avenue to solve this problem is to introduce new fields to the action to eat up the large vacuum energy density 
$( \rho_{vac} \sim 10^{72}~ \text{GeV}^4)$ and reduce it to the observed dark energy density $ \leq (10^{-47}~ \text{GeV}^4)$. In this context, we model the universe as a 3-brane embedded in the bulk of $\N=2$ $\D=5$ ungauged supergravity (SUGRA) which results by compatifying  $\N=1$ $\D=11$ supergravity over a Calabi- Yau threefold ($\M$) \cite{Ceresole:2000jd, Celi:2003qk}. 

%\vspace{1cm}

On the brane, an effective energy density is produced which is adequate for the observations. Also, the cosmic evolution of the brane universe corresponds to the most recent observed parameters of the $\Lambda$CDM model \cite{Peebles:1984, Turner:2022}, due to the strong correlation between the flow velocity of the complex structure moduli and the brane’s scale factor. The bulk shows a decelerated contraction due to the effect of its negative and large energy density. The  6-extra dimensions of $\D=5$ supergravity are compactified on a scale $(\sim 10^{-12} ~\text{m}) $ according to the framework of Arkani-Hamed, Dimopoulos, and Dvali (ADD) \cite{ Arkani:1998}, so the electroweak hierarchy problem has been implicitly explained on the brane due to the existence of large extra dimensions compared to millimeters and by interpolating the electroweak scale as the fundamental scale of nature.
All these consists with the experiment since there are no pieces of evidence for extra dimensions that evolve with time like our universe, while any deviations from Newton's gravity at sub-millimeters scales or less may be found by future experiments. It is forth mentioning that the electroweak hierarchy problem has been investigated in details through $N=2$ $\D=5$ supergravity theory in \cite{Salem:2023kzm}. 
There have been many attempts to solve the cosmological constant problem or the cosmological hierarchy like \cite{Bertolami:2010, Rasouli:2020} and within the brane-worlds scenarios, \cite{Burgess:1999qw, Iglesias:2003di, Kang:2019cbd, Chiou-Lahanas:2013eoa, Kang:2019ymk}. For instance, in
\cite{Canestaro:2013xsa} the role of the universal hypermultiplet of $\N=2$ $\D=5$ SUGRA,
specifically the dilaton, has been investigated in reducing the value of the 3-brane cosmological constant in the modified Friedmann equations, which has been considered as the one related to the vacuum energy. 
Here, however, we study the role of the Calabi-Yau threefold complex structure moduli and the metric of the K\"{a}hler space of $\M$ . This will shed light on the topology of the Calabi-Yau manifold itself and how it can drive the dynamics of both the brane-world and the bulk.
Also in a previous paper \cite{Emam:2023idq}, we have shown how the problem can be solved by 
interpreting the late-time acceleration of the universe by topological effects, where the brane-universe had a vanishing cosmological constant, and the brane filled only by matter and radiation. Also, the dependence of the moudli on the fifth dimension hasn't been taken into account. 
Nevertheless, it is dangerous to interpret the cosmological constant problem in the absence of dark energy, because that makes 
the current matter energy density of the brane-universe around $\Omega_{m0} \sim 0.99 $. This means the current dark matter energy density
$\Omega_{Dm0} \sim 0.99 $, since the current baryonic matter energy density $\Omega_{Bm0} \sim 0.04$, while the increasing of 
$\Omega_{Dm0}$ may has cosmological impacts in the early universe like the Baryon Acoustic Oscillations \cite{Menote:2021}.

In this paper, the matter sector of the theory is comprised of four scalar fields and their superpartners; they are known as the universal hypermultiplets ( $\sigma, \phi, \zeta^0,\tilde{\zeta}_0$ ); the field  $\phi$ is called the universal axion, and the field $\sigma$ known as the dilaton. The rest of the hypermultiplets 
scalars are $\left(z^i, z^{\bar i}, \zeta^i, \tilde \zeta_i: i=1,\ldots, \htwo\right)$, where the $z$'s are the complex structure moduli of $\M$ and $\htwo$ is the Hodge number determining the dimensions of the complex structure space $\M_C$ of the Calabi-Yau's manifold ($\M$). The ‘bar’ over an index denotes complex conjugation. The fields $\left(\zeta^I, \tilde\zeta_I: I=0,\ldots,\htwo\right)$ are called axions. While the vector multiplets of the theory trivially decouple from the hypermultiplets and can be set to zero. We consider the dependence of the hypermultiplets on both time and the fifth extra dimension.

The paper is organized as follows, in section \RNum{2} we introduce the model, the action, the fields equations of motions, the gravitini and the hyperini supersymmetry (SUSY) transformations, the energy momentum tensor
and the yielded Bogomol’nyi- Prasad- Sommerfield (BPS) condition. In section \RNum{3} we introduce the metric of the model, and the modified Friedmann equations, and show how the cosmological constant problem can be solved. In section \RNum{4} we calculate the size of the extra dimensions and investigate the time evolution of the bulk's scale factor. Also, we show how consequently there is no electroweak hierarchy in the model. In section \RNum{5} we wrap our results.

%e
\section{The model and its BPS condition} 

The axions are defined as components of the symplectic vector
\be\label{DefOfSympVect}
   \left| \Xi  \right\rangle  = \left( {\begin{array}{*{20}c}
   {\,\,\,\,\,\zeta ^I }  \\
   -{\tilde \zeta _I }  \\
    \end{array}} \right),
\ee
such that the symplectic scalar product is defined by, for example,
\be
    \left\langle {{\Xi }}
 \mathrel{\left | {\vphantom {{\Xi } \Xi }}
 \right. \kern-\nulldelimiterspace}
 {\Xi } \right\rangle   = \zeta^I \tilde \zeta_I  - \tilde \zeta_I
 \zeta^I.\label{DefOfSympScalarProduct}
\ee

A transformation in symplectic space can be defined by
\be
 \left\langle {d\Xi } \right|\mathop {\bf\Lambda} \limits_ \wedge  \left| {\star d\Xi } \right\rangle
  = 2\left\langle {{d\Xi }}
 \mathrel{\left | {\vphantom {{d\Xi } V}}
 \right. \kern-\nulldelimiterspace}
 {V} \right\rangle \mathop {}\limits_ \wedge  \left\langle {{\bar V}}
 \mathrel{\left | {\vphantom {{\bar V} {\star d\Xi }}}
 \right. \kern-\nulldelimiterspace}
 {{\star d\Xi }} \right\rangle  + 2G^{i\bar j} \left\langle {{d\Xi }}
 \mathrel{\left | {\vphantom {{d\Xi } {U_{\bar j} }}}
 \right. \kern-\nulldelimiterspace}
 {{U_{\bar j} }} \right\rangle \mathop {}\limits_ \wedge  \left\langle {{U_i }}
 \mathrel{\left | {\vphantom {{U_i } {\star d\Xi }}}
 \right. \kern-\nulldelimiterspace}
 {{\star d\Xi }} \right\rangle  - i\left\langle {d\Xi } \right.\mathop |\limits_ \wedge  \left. {\star d\Xi } \right\rangle,\label{DefOfRotInSympSpace}
\ee
where $d$ is the spacetime exterior derivative, $\star$ is the five dimensional Hodge duality operator, and $G_{i\bar j}$ is a special K\"{a}hler metric on $\M_C$. The symplectic basis vectors $\left| V \right\rangle $, $\left| {U_i } \right\rangle $ and their complex conjugates are defined by
\be
    \left| V \right\rangle  = e^{\frac{\K}{2}} \left( {\begin{array}{*{20}c}
   {Z^I }  \\
   {F_I }  \\
    \end{array}} \right),\,\,\,\,\,\,\,\,\,\,\,\,\,\,\,\left| {\bar V} \right\rangle  = e^{\frac{\K}{2}} \left( {\begin{array}{*{20}c}
   {\bar Z^I }  \\
   {\bar F_I }  \\
    \end{array}} \right)\label{DefOfVAndVBar}
\ee

\noindent where $\K$ is the K\"{a}hler potential on $\M_C$, $\left( {Z,F} \right)$ are the periods of the Calabi-Yau's holomorphic volume form, and

\bea
    \left| {U_i } \right\rangle  &=& \left| \nabla _i V
    \right\rangle=\left|\left[ {\partial _i  + \frac{1}{2}\left( {\partial _i \K} \right)} \right] V \right\rangle \nonumber\\
    \left| {U_{\bar i} } \right\rangle  &=& \left|\nabla _{\bar i}  {\bar V} \right\rangle=\left|\left[ {\partial _{\bar i}  + \frac{1}{2}\left( {\partial _{\bar i} \K} \right)} \right] {\bar V}
    \right\rangle\label{DefOfUAndUBar}
\eea
where the derivatives are with respect to the moduli $\left(z^i, z^{\bar i}\right)$. In this language, the bosonic part of the action is given by:
\bea
    S_5  &=& \int\limits_5 {\left[ {R\star \mathbf{1} - \frac{1}{2}d\sigma \wedge\star d\sigma  - G_{i\bar j} dz^i \wedge\star dz^{\bar j} } \right.}  + e^\sigma   \left\langle {d\Xi } \right|\mathop {\bf\Lambda} \limits_ \wedge  \left| {\star d\Xi } \right\rangle\nonumber\\
    & &\left. {\quad\quad\quad\quad\quad\quad\quad\quad\quad\quad\quad\quad\quad - \frac{1}{2} e^{2\sigma } \left[ {d\phi + \left\langle {\Xi } \mathrel{\left | {\vphantom {\Xi  {d\Xi }}} \right. \kern-\nulldelimiterspace} {{d\Xi }}    \right\rangle} \right] \wedge \star\left[ {d\phi + \left\langle {\Xi } \mathrel{\left | {\vphantom {\Xi  {d\Xi }}} \right. \kern-\nulldelimiterspace} {{d\Xi }}    \right\rangle} \right] } \right].\label{action}
\eea%

The usual $\delta S = 0$ gives the following field equations for the hypermultiplets scalar fields:

\bea
    \left( {\Delta \sigma } \right)\star \mathbf{1} + e^\sigma   \left\langle {d\Xi } \right|\mathop {\bf\Lambda} \limits_ \wedge  \left| {\star d\Xi } \right\rangle -   e^{2\sigma }\left[ {d\phi + \left\langle {\Xi } \mathrel{\left | {\vphantom {\Xi  {d\Xi }}} \right. \kern-\nulldelimiterspace} {{d\Xi }}    \right\rangle} \right]\wedge\star\left[ {d\phi + \left\langle {\Xi } \mathrel{\left | {\vphantom {\Xi  {d\Xi }}} \right. \kern-\nulldelimiterspace} {{d\Xi }}    \right\rangle} \right] &=& 0\label{DilatonEOM}\\
    \left( {\Delta z^i } \right)\star \mathbf{1} + \Gamma _{jk}^i dz^j  \wedge \star dz^k  + \frac{1}{2}e^\sigma  G^{i\bar j}  {\partial _{\bar j} \left\langle {d\Xi } \right|\mathop {\bf\Lambda} \limits_ \wedge  \left| {\star d\Xi } \right\rangle} &=& 0 \nonumber\\
    \left( {\Delta z^{\bar i} } \right)\star \mathbf{1} + \Gamma _{\bar j\bar k}^{\bar i} dz^{\bar j}  \wedge \star dz^{\bar k}  + \frac{1}{2}e^\sigma  G^{\bar ij}  {\partial _j \left\langle {d\Xi } \right|\mathop {\bf\Lambda} \limits_ \wedge  \left| {\star d\Xi } \right\rangle}  &=& 0\label{ZZBarEOM} \\
    d^{\dag} \left\{ {e^\sigma  \left| {{\bf\Lambda} d\Xi } \right\rangle  - e^{2\sigma } \left[ {d\phi + \left\langle {\Xi }
    \mathrel{\left | {\vphantom {\Xi  {d\Xi }}}\right. \kern-\nulldelimiterspace} {{d\Xi }} \right\rangle } \right]\left| \Xi  \right\rangle } \right\} &=& 0\label{AxionsEOM}\\
    d^{\dag} \left[ {e^{2\sigma } d\phi + e^{2\sigma } \left\langle {\Xi } \mathrel{\left | {\vphantom {\Xi  {d\Xi }}} \right. \kern-\nulldelimiterspace} {{d\Xi }}    \right\rangle} \right] &=&    0\label{aEOM}
\eea
where $d^\dagger$ is the $D=5$ adjoint exterior derivative, $\Delta$ is the Laplace-de Rahm operator and $\Gamma _{jk}^i$ is a connection on $\M_C$. The full action is symmetric under the following SUSY transformations:
\bea
 \delta _\epsilon  \psi ^1  &=& D \epsilon _1  + \frac{1}{4}\left\{ {i {e^{\sigma } \left[ {d\phi + \left\langle {\Xi }
 \mathrel{\left | {\vphantom {\Xi  {d\Xi }}}
 \right. \kern-\nulldelimiterspace} {{d\Xi }} \right\rangle } \right]}- Y} \right\}\epsilon _1  - e^{\frac{\sigma }{2}} \left\langle {{\bar V}}
 \mathrel{\left | {\vphantom {{\bar V} {d\Xi }}} \right. \kern-\nulldelimiterspace} {{d\Xi }} \right\rangle\epsilon _2  \nonumber\\
 \delta _\epsilon  \psi ^2  &=& D \epsilon _2  - \frac{1}{4}\left\{ {i {e^{\sigma } \left[ {d\phi + \left\langle {\Xi }
 \mathrel{\left | {\vphantom {\Xi  {d\Xi }}} \right. \kern-\nulldelimiterspace}
 {{d\Xi }} \right\rangle } \right]}- Y} \right\}\epsilon _2  + e^{\frac{\sigma }{2}} \left\langle {V}
 \mathrel{\left | {\vphantom {V {d\Xi }}} \right. \kern-\nulldelimiterspace} {{d\Xi }} \right\rangle \epsilon _1,  \label{SUSYGraviton}
\eea
\bea
  \delta _\epsilon  \xi _1^0  &=& e^{\frac{\sigma }{2}} \left\langle {V}
    \mathrel{\left | {\vphantom {V {\partial _\mu  \Xi }}} \right. \kern-\nulldelimiterspace} {{\partial _\mu  \Xi }} \right\rangle  \Gamma ^\mu  \epsilon _1  - \left\{ {\frac{1}{2}\left( {\partial _\mu  \sigma } \right) - \frac{i}{2} e^{\sigma } \left[ {\left(\partial _\mu \phi\right) + \left\langle {\Xi }
    \mathrel{\left | {\vphantom {\Xi  {\partial _\mu \Xi }}} \right. \kern-\nulldelimiterspace}
    {{\partial _\mu \Xi }} \right\rangle } \right]} \right\}\Gamma ^\mu  \epsilon _2  \nonumber\\
     \delta _\epsilon  \xi _2^0  &=& e^{\frac{\sigma }{2}} \left\langle {{\bar V}}
    \mathrel{\left | {\vphantom {{\bar V} {\partial _\mu  \Xi }}} \right. \kern-\nulldelimiterspace} {{\partial _\mu  \Xi }} \right\rangle \Gamma ^\mu  \epsilon _2  + \left\{ {\frac{1}{2}\left( {\partial _\mu  \sigma } \right) + \frac{i}{2} e^{\sigma } \left[ {\left(\partial _\mu \phi\right) + \left\langle {\Xi }
    \mathrel{\left | {\vphantom {\Xi  {\partial _\mu \Xi }}} \right. \kern-\nulldelimiterspace}
    {{\partial _\mu \Xi }} \right\rangle } \right]} \right\}\Gamma ^\mu  \epsilon
     _1,\label{SUSYHyperon1}
\eea
and
\bea
     \delta _\epsilon  \xi _1^{\hat i}  &=& e^{\frac{\sigma }{2}} e^{\hat ij} \left\langle {{U_j }}
    \mathrel{\left | {\vphantom {{U_j } {\partial _\mu  \Xi }}} \right. \kern-\nulldelimiterspace} {{\partial _\mu  \Xi }} \right\rangle \Gamma ^\mu  \epsilon _1  - e_{\,\,\,\bar j}^{\hat i} \left( {\partial _\mu  z^{\bar j} } \right)\Gamma ^\mu  \epsilon _2  \nonumber\\
     \delta _\epsilon  \xi _2^{\hat i}  &=& e^{\frac{\sigma }{2}} e^{\hat i\bar j} \left\langle {{U_{\bar j} }}
    \mathrel{\left | {\vphantom {{U_{\bar j} } {\partial _\mu  \Xi }}} \right. \kern-\nulldelimiterspace} {{\partial _\mu  \Xi }} \right\rangle \Gamma ^\mu  \epsilon _2  + e_{\,\,\,j}^{\hat i} \left( {\partial _\mu  z^j } \right)\Gamma ^\mu  \epsilon    _1,\label{SUSYHyperon2}
\eea
where $\left(\psi ^1, \psi ^2\right)$ are the two gravitini and $\left(\xi _1^I, \xi _2^I\right)$ are the hyperini. The quantity $Y$ is defined by:
\begin{equation}
    Y   = \frac{{\bar Z^I N_{IJ}  {d  Z^J }  -
    Z^I N_{IJ}  {d  \bar Z^J } }}{{\bar Z^I N_{IJ} Z^J
    }},\label{DefOfY}
\end{equation}

where $N_{IJ}  = \mathfrak{Im} \left({\partial_IF_J } \right)$. The $e$'s are the beins of the special K\"{a}hler metric $G_{i\bar j}$, the $\epsilon$'s are the five-dimensional $\N=2$ SUSY spinors and the $\Gamma$'s are the usual Dirac matrices. The covariant derivative $D$ is defined by the usual $D=dx^\mu\left( \partial _\mu   + \frac{1}{4}\omega _\mu^{\,\,\,\,\hat \mu\hat \nu} \Gamma _{\hat \mu\hat \nu}\right)\label{DefOfCovDerivative}$, where the $\omega$'s are the spin connections and the hatted indices are frame indices in a flat tangent space. 

The total energy stress tensor is given by: 
\be
T_{\mu\nu} = T_{\mu\nu}^{Bulk} + T_{\mu\nu}^{3b},
\ee
where $\mu,\nu=0,1,…3$, $T_{\mu\nu}^{Bulk}$ and $T_{\mu\nu}^{3b}$ are the bulk and the 3- brane energy stress tensors, respectively. $T_{\mu\nu}^{3b}$ is given by the usual perfect fluid energy stress tensor:
\be
T_{\mu\nu}^{3b}=  (p+\rho) U_\mu U_\nu + g_{\mu\nu} p,
\ee
where $T^{3b}_{tt} = \rho, \text{and}~ T^{3b}_{rr} = e^{2\beta}$, $T_{yt} =0$.  
\bea
\nonumber T^{Bulk} &=&  -\frac{1}{2} \partial_\mu \sigma \partial_\nu \sigma +  \frac{1}{4} g_{\mu\nu} \partial_\alpha \sigma \partial^\alpha \sigma  + e^\sigma [\partial_\mu \Xi|\Lambda|\partial_\nu \Xi ] - \frac{1}{2} g_{\mu\nu}  e^\sigma[\partial_\alpha \Xi|\Lambda|\partial^\alpha \Xi ] \\ \nonumber    
 &-& \frac{e^{2\sigma}}{2} [\partial_\mu \phi + \langle \Xi|\partial_\mu \Xi\rangle ] ~ [ \partial_\nu \phi+ \langle \Xi|\partial_\nu \Xi\rangle ] + \frac{1}{4} g_{\mu\nu} e^{2\sigma} [\partial_\alpha \phi + \langle \Xi|\partial_\alpha \Xi\rangle ] ~ [ \partial^\alpha \phi+ \langle \Xi|\partial^\alpha \Xi\rangle ]
\\ \nonumber &-& G_{i\bar{j}} \partial_\mu z^i \partial_\nu z^{\bar{j}} + \frac{1}{2} g_{\mu\nu} G_{i\bar{j}} \partial_\alpha z^i \partial^\alpha z^{\bar{j}}  .
\eea
\bea
\nonumber T_{yy} &=&  -\frac{1}{2} \partial_\mu \sigma \partial_\nu \sigma +  \frac{1}{4} g_{\mu\nu} \partial_\alpha \sigma \partial^\alpha \sigma  + e^\sigma [\partial_\mu \Xi|\mathbf{\Lambda}|\partial_\nu \Xi ] - \frac{1}{2} g_{\mu\nu}  e^\sigma[\partial_\alpha \Xi|\mathbf{\Lambda}|\partial^\alpha \Xi ] \\ \nonumber    
 &-& \frac{e^{2\sigma}}{2} [\partial_\mu \phi + \langle \Xi|\partial_\mu \Xi\rangle ] ~ [ \partial_\nu \phi+ \langle \Xi|\partial_\nu \Xi\rangle ] + \frac{1}{4} g_{\mu\nu} e^{2\sigma} [\partial_\alpha \phi + \langle \Xi|\partial_\alpha \Xi\rangle ] ~ [ \partial^\alpha \phi+ \langle \Xi|\partial^\alpha \Xi\rangle ]
\\ \nonumber &-& G_{i\bar{j}} \partial_\mu z^i \partial_\nu z^{\bar{j}} + \frac{1}{2} g_{\mu\nu} G_{i\bar{j}} \partial_\alpha z^i \partial^\alpha z^{\bar{j}}  .
\eea 
To have a BPS solution for the theory that preserves supersymmetry, the fermions SUSY transformations (\ref{SUSYHyperon1},\ref{SUSYHyperon2}) should vanish, means that all the fermions in the theory vanish and we get a bosonic solution. Worthy to mention that in supergravity supersymmetry is half broken. That yields the BPS condition: 
\be e^\sigma \langle \Xi|\underset{\Lambda}{\mathbf{\Lambda}}| \star d\Xi \rangle
 = \frac{1}{2} d\sigma \wedge \star d\sigma + \frac{1}{2} e^{2\sigma} [d\phi + \langle\Xi|~d\Xi \rangle] \wedge \star [d\phi + \langle \Xi|~d\Xi\rangle] + 2 G_{i\bar{j}} dz^i \wedge \star dz^{\bar{j}},\ee
as outlined in details in \cite{Emam:2020oyb}. Substitute into the energy stress tensors, they are simplified to 
\bea
\nonumber
T_{\mu\nu}^{Bulk} &=&  G_{i\bar{j}} \partial_\mu z^i \partial_\nu z^{\bar{j}} - \frac{1}{2} g_{\mu\nu} G_{i\bar{j}} \partial_\alpha z^i \partial^\alpha z^{\bar{j}}, \\
T_{yy} &=& G_{i\bar{j}} \partial_y z^i \partial_y  z^{\bar{j}} - \frac{1}{2} g_{yy} G_{i\bar{j}} \partial_\alpha z^i \partial^\alpha z^{\bar{j}}.
\eea

\section{The evolution of the brane-universe and the late-time acceleration}

We consider a flat Friedman-Lemaître- Robertson-Walker (FLRW) metric embedded in a five dimensional bulk
\be%
ds^2= -  dt^2 + a^2(t) (dr^2 + r^2 d \Omega^2 ) + b^2\left( t \right) dy^2,
\ee%
where $d\Omega^2=d\theta^2+\sin^2\theta~ d\phi^2$, $a \left( t \right)$ is the 3-brane's scale factor, and $b \left( t \right)$ is  the scale factor of the bulk.
%…………………………
Solving the Einstein's equations:
\bea
\nonumber G_{\mu\nu}+\Lambda g_{\mu\nu} & =& T_{\mu\nu}^{Bulk} + T_{\mu\nu}^{3b} \\
G_{yy}+\tilde{\Lambda} g_{yy} & =& T_{yy}, ~~~~~~~~ G_{yt}= T_{yt}=0,
\eea
gives:
\bea
 \left[ {\left( {\frac{{\dot a}}{a}} \right)^2  + \left( {\frac{{\dot a}}{a}} \right)\left( {\frac{{\dot b}}{b}} \right)} \right] &=& G_{i\bar j} \dot z^i \dot z^{\bar j}  + \frac{8 \pi G}{3} \left ( \rho_m + \rho_r + \rho_{vac} \right) + G_{i\bar{j}} ~z^{i'} ~ z^{\bar{j}'}  \nonumber\\
 2\frac{{\ddot a}}{a} + \left( {\frac{{\dot a}}{a}} \right)^2  + \frac{{\ddot b}}{b} + 2\left( {\frac{{\dot a}}{a}} \right)\left( {\frac{{\dot b}}{b}} \right) &=&  \frac{8 \pi G}{3}  \left (  \rho_{vac}  - \rho_r \right)  - G_{i\bar j} \dot z^i \dot z^{\bar j} + G_{i\bar{j}} ~z^{i'} ~ z^{\bar{j}'}\nonumber\\
 \left[ {\frac{{\ddot a}}{a} + \left( {\frac{{\dot a}}{a}} \right)^2 } \right] &=&  \tilde\Lambda  - G_{i\bar j} \dot z^i \dot z^{\bar j} - G_{i\bar{j}} ~z^{i'} ~ z^{\bar{j}'} ,
\eea

\begin{figure}[t]
  \begin{subfigure}[t]{.5\linewidth}
    \centering
    \includegraphics[width=1\columnwidth]{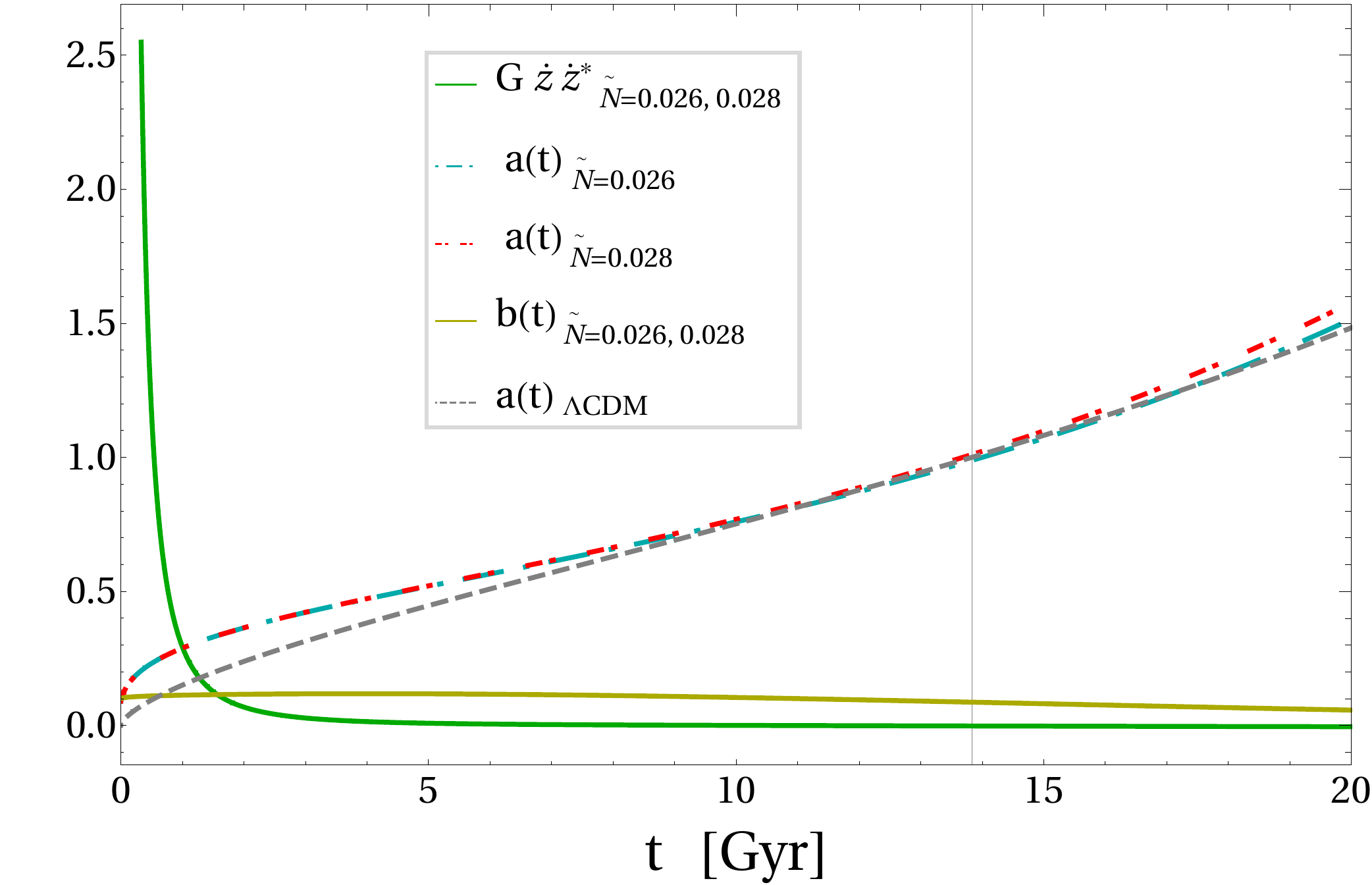}
    \caption{The time evolution of the scale factor of $\Lambda$CDM is represented by the gray dotted curve, the brane's scale factor $a$ is represented by the cyan dashed curve and the red dashed curve for $\tilde{N}=0.026$ and $0.028$, respectively, the bulk's scale factor $b$ by the green curve, and $\left| {G_{i\bar j} \dot z^i \dot z^{\bar j}} \right|$ by the light green curve. The initial conditions are: $a(0)= 0.09,~ \dot{a}(0)=1, b(0)= 0.01, \dot{b}(0)=0.01$. The vertical grid line is at $t_0= 13.842~ [\text{Gyr}]$.}
    \label{abz02}
  \end{subfigure}
\qquad 
  \begin{subfigure}[t]{.5\linewidth}
    \centering
    \includegraphics[width=1\columnwidth]{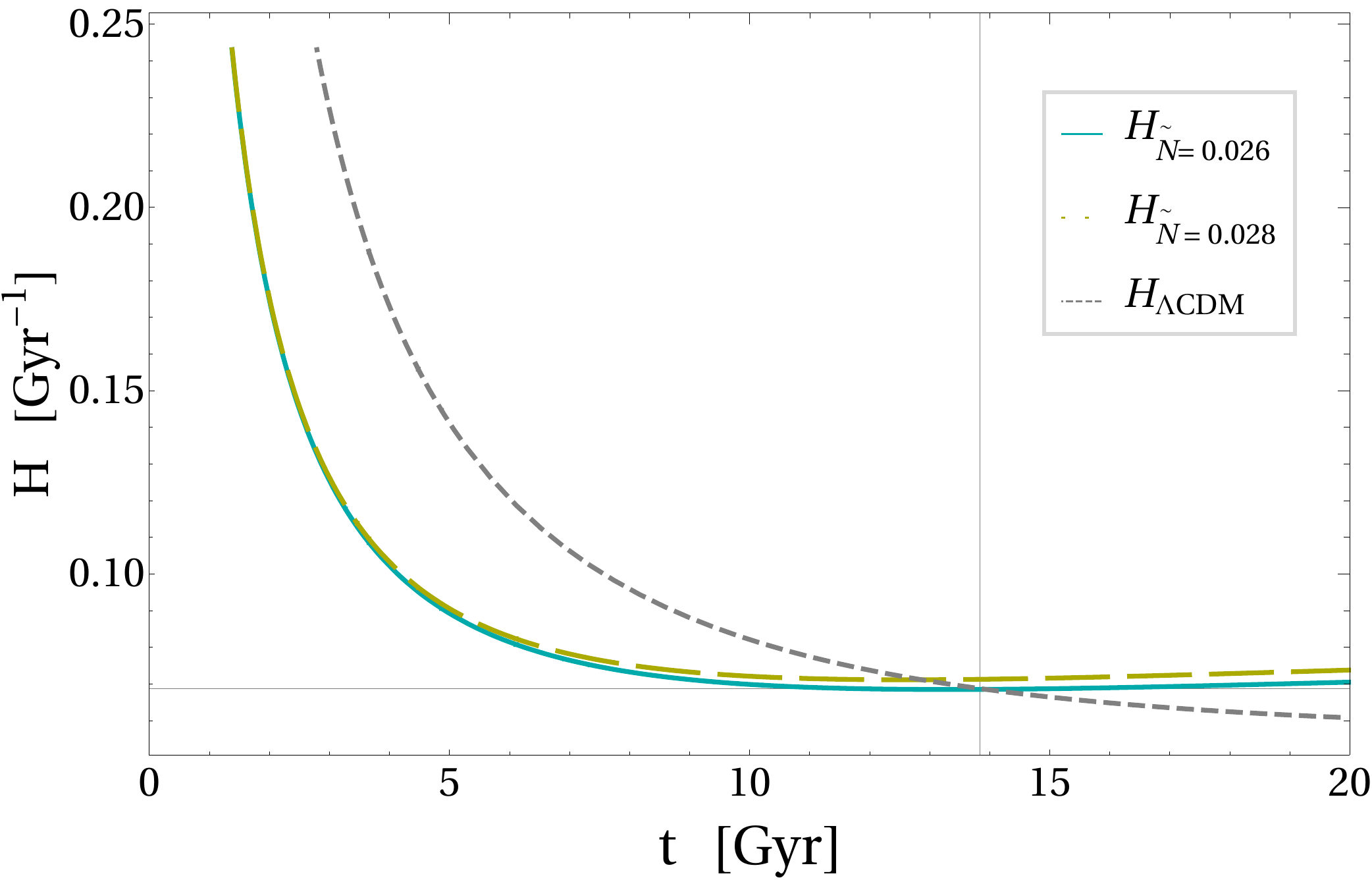}
    \caption{The Hubble parameter of $\Lambda$CDM in the gray dotted curve, and
the Hubble parameter of the brane in the cyan solid curve and the green dashed curve for $\tilde{N}=0.026$ and $0.028$, respectively are plotted versus time. The horizontal grid line is at $H_0= 0.0689751 ~[\text{Gyr}^{-1}]$.}
    \label{hub}
  \end{subfigure}
\caption{}
\end{figure}

The norm of the moduli flow with respect to $y$ can be added to $\rho_{vac}$ to reduce it to the observed dark energy density, such that the effective observed dark energy density is given by $\rho_{eff} \equiv \rho_\Lambda = \rho_{vac} + \frac{3} {8 \pi G} G_{i\bar{j}} ~z^{i'} ~ z^{\bar{j}'} $. This means $|G_{i\bar{j}} ~z^{i'} ~ z^{\bar{j}'}|$ should be around $\sim  10^{83}~ [\text{sec}^{-2}] $ to reduce $\rho_{vac}$  to around $\rho_{\Lambda} = 10^{-47}~ GeV^4 $. In natural units $|G_{i\bar{j}} ~z^{i'} ~ z^{\bar{j}'}|$ is around $ \sim 10^{35}~ GeV^2$. This value is theoretically valid since it is less than $\mathcal{O} (2)$ of $M_{pl}$. Whilst $\tilde\Lambda$ in the third field equations should to be fine-tuned to cancel the large value of $G_{i\bar{j}} ~z^{i'} ~ z^{\bar{j}'}$, for instance, take $\tilde\Lambda \sim (\tilde N- \frac{8\pi G}{3} \rho_{vac} )\equiv (\tilde N- G_{i\bar{j}} ~z^{i'} ~ z^{\bar{j}'} )$. The tuning of $\tilde\Lambda$ can be justified by assuming a high distribution of quantum fields in the bulk which is physically reasonable. Then the field equations become;
\bea
 \left[ {\left( {\frac{{\dot a}}{a}} \right)^2  + \left( {\frac{{\dot a}}{a}} \right)\left( {\frac{{\dot b}}{b}} \right)} \right] &=& G_{i\bar j} \dot z^i \dot z^{\bar j}  + \frac{8 \pi G}{3} \left ( \rho_m + \rho_r + \rho_\Lambda \right) \nonumber\\
 2\frac{{\ddot a}}{a} + \left( {\frac{{\dot a}}{a}} \right)^2  + \frac{{\ddot b}}{b} + 2\left( {\frac{{\dot a}}{a}} \right)\left( {\frac{{\dot b}}{b}} \right) &=&  \frac{8 \pi G}{3}  \left (  \rho_\Lambda  - \rho_r \right)  - G_{i\bar j} \dot z^i \dot z^{\bar j}  \nonumber\\
 \left[ {\frac{{\ddot a}}{a} + \left( {\frac{{\dot a}}{a}} \right)^2 } \right] &=&  \tilde N  - G_{i\bar j} \dot z^i \dot z^{\bar j} ,
\eea
where $\rho_m$, $\rho_r$ are the matter and the radiation energy densities, respectively. Or in terms of the density parameters;
\bea
 \left[ {\left( {\frac{{\dot a}}{a}} \right)^2  + \left( {\frac{{\dot a}}{a}} \right)\left( {\frac{{\dot b}}{b}} \right)} \right] &=& G_{i\bar j} \dot z^i \dot z^{\bar j}  + H_0^2 \left ( \frac{{\Omega_{m0} }}{{a^3 }} +  \frac{\Omega_{r0} }{{a^4 }} + \Omega_{\Lambda0} \right) \nonumber\\
 2\frac{{\ddot a}}{a} + \left( {\frac{{\dot a}}{a}} \right)^2  + \frac{{\ddot b}}{b} + 2\left( {\frac{{\dot a}}{a}} \right)\left( {\frac{{\dot b}}{b}} \right) &=&  H_0^2 \Omega_{\Lambda0}  - H_0^2 \frac{\Omega_{r0}}{a^4} - G_{i\bar j} \dot z^i \dot z^{\bar j}  \nonumber\\
 \left[ {\frac{{\ddot a}}{a} + \left( {\frac{{\dot a}}{a}} \right)^2 } \right] &=&  \tilde N - G_{i\bar j} \dot z^i \dot z^{\bar j}.
\label{DPE}
\eea
Or in terms of the brane's and the bulk's Hubble parameters $H=\frac{\dot{a}}{a}$ and $\tilde{H}=\frac{\dot{b}}{b}$, respectively;
\bea
  H^2  +  H  \tilde{H} &=& G_{i\bar j} \dot z^i \dot z^{\bar j}  + \frac{8 \pi G}{3} \left ( \rho_m + \rho_r + \rho_\Lambda \right) \nonumber\\
 2 \dot{H} + 3 H^2 + \dot{\tilde{H}} + \tilde{H}^2 + 2 H \tilde{H}   &=&  \frac{8 \pi G}{3}  \left (  \rho_\Lambda  - \rho_r \right)  - G_{i\bar j} \dot z^i \dot z^{\bar j}  \nonumber\\
 \left[ \dot{H} + 2 H^2 \right] &=&  \tilde N  - G_{i\bar j} \dot z^i \dot z^{\bar j} ,
\eea
We solve equations (\ref{DPE}) numerically to find $a, b$ and $G_{i\bar j} \dot z^i \dot z^{\bar j}$ for different values of $\tilde{N}$.
The current values of the density parameters of the matter, radiation, and dark energy  corresponding to the values of $\Lambda$CDM model are given by: $\Omega_{m0} = 0.3111$, $\Omega_{r0}= 8.2 \times 10^{-5}$, and $\Omega_{\Lambda 0} = 0.6889$. 
The current value of the Hubble parameter is given by $H_0= 0.0686751 ~[\text{Gyr}^{-1}] \sim 2.176 \times 10^{-18}~ [\text{sec}^{-1}] $. The present time of the universe is $t_0= 13.842~ [\text{Gyr}] =1 \times 10^{10}~ [\text{Year}]$ \cite{Planck:2018}. 
Fig. (\ref{abz02}) shows
the time evolution of the scale factor of $\Lambda$CDM model represented by the dotted gray curve, 
the brane's scale factor for $\tilde{N
}= 0.026$ by the cyan dashed curve and for $\tilde{N}= 0.028$ by the red dashed curve, the bulk's scale factor by the yellow curve for $\tilde{N}= 0.026~ \text{and}~ 0.028$, and the moduli velocity norm by the green curve for $\tilde{N}= 0.026~ \text{and}~ 0.028$. This plot indicates the strong relation between $\left| {G_{i\bar j} \dot z^i \dot z^{\bar j}} \right|$ and
the brane's scale factor. To fit $\Lambda$CDM observations  $\tilde{N}$ should be around $0.02 < \tilde{N} <0.03$ $[\text{Gyr}^{-2}]$.      
%%%%%%%%
In figure (\ref{hub}) the time evolution of $\Lambda$CDM's Hubble parameter ($H_{\Lambda CDM}$), and the Hubble parameter
for the brane- universe for $\tilde{N}= 0.026$ in the cyan solid curve and for $\tilde{N}= 0.028$ in the yellow dashed curve are plotted versus time. The Brane's Hubble parameters evolve slightly less than that of $\Lambda$CDM, but 
$H_{\tilde{N}= 0.028}$ meets $H_{\Lambda CDM}$ at the current era of the universe ($t_0$). 
%--------------------------------------------------------
\vspace{1cm}

\begin{figure}[t]
  \begin{subfigure}[t]{.48\linewidth}
    \centering
    \includegraphics[width=1\columnwidth]{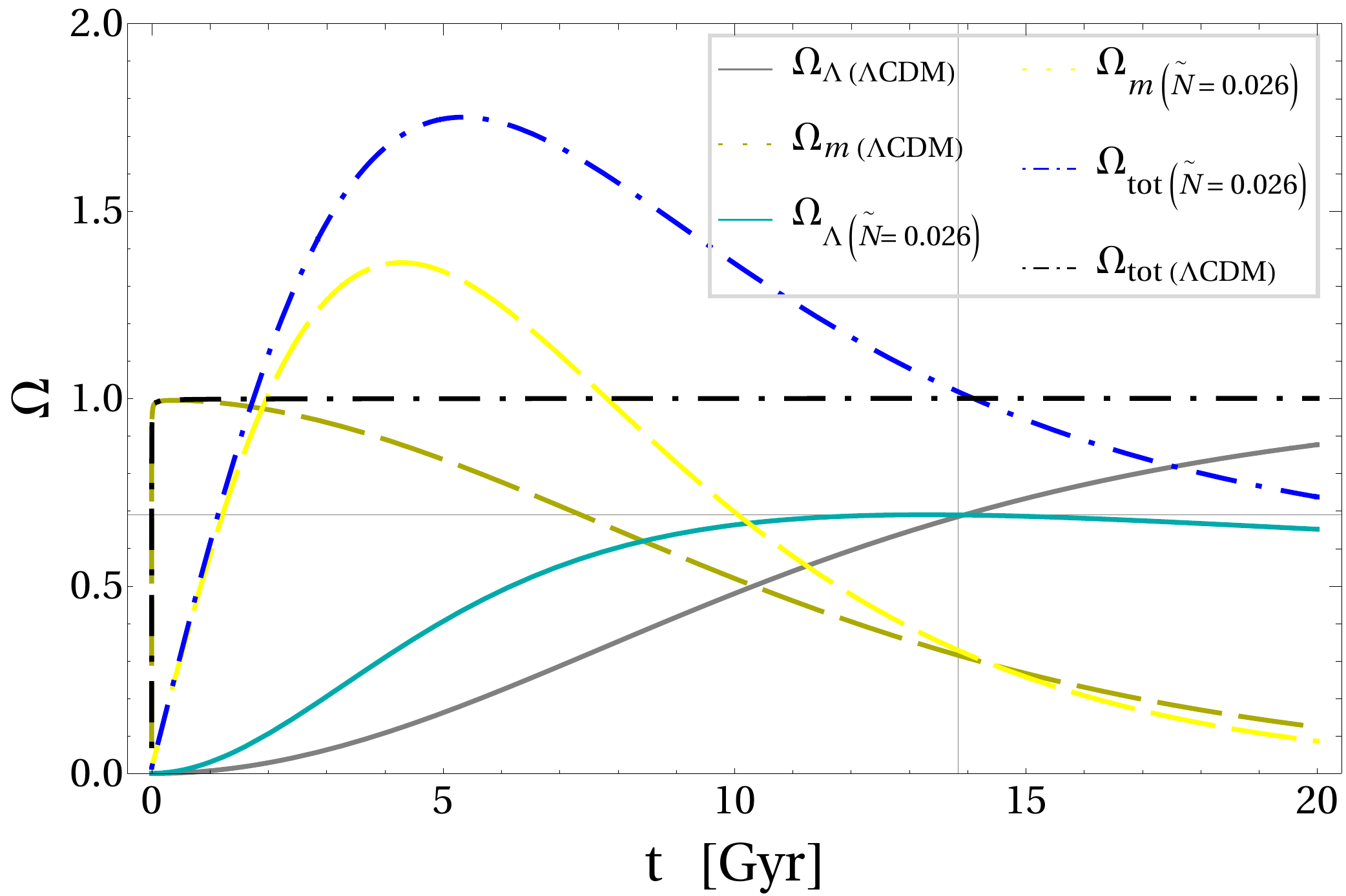}
    \caption{The time evolution of 
the matter and energy density parameters of $\Lambda$CDM are
plotted in the green dashed curve and the gray solid curve, respectively. The total $\Lambda$CDM's density parameter is plotted by the black dashed curve. The matter and energy density parameters of the brane are
plotted by the yellow dashed curve and the solid cyan curve, respectively. The total brane's density parameter is plotted by the blue dashed curve.  
}
    \label{allDpar1}
  \end{subfigure} 
\qquad 
  \begin{subfigure}[t]{.51\linewidth}
    \centering
    \includegraphics[width=1\columnwidth]{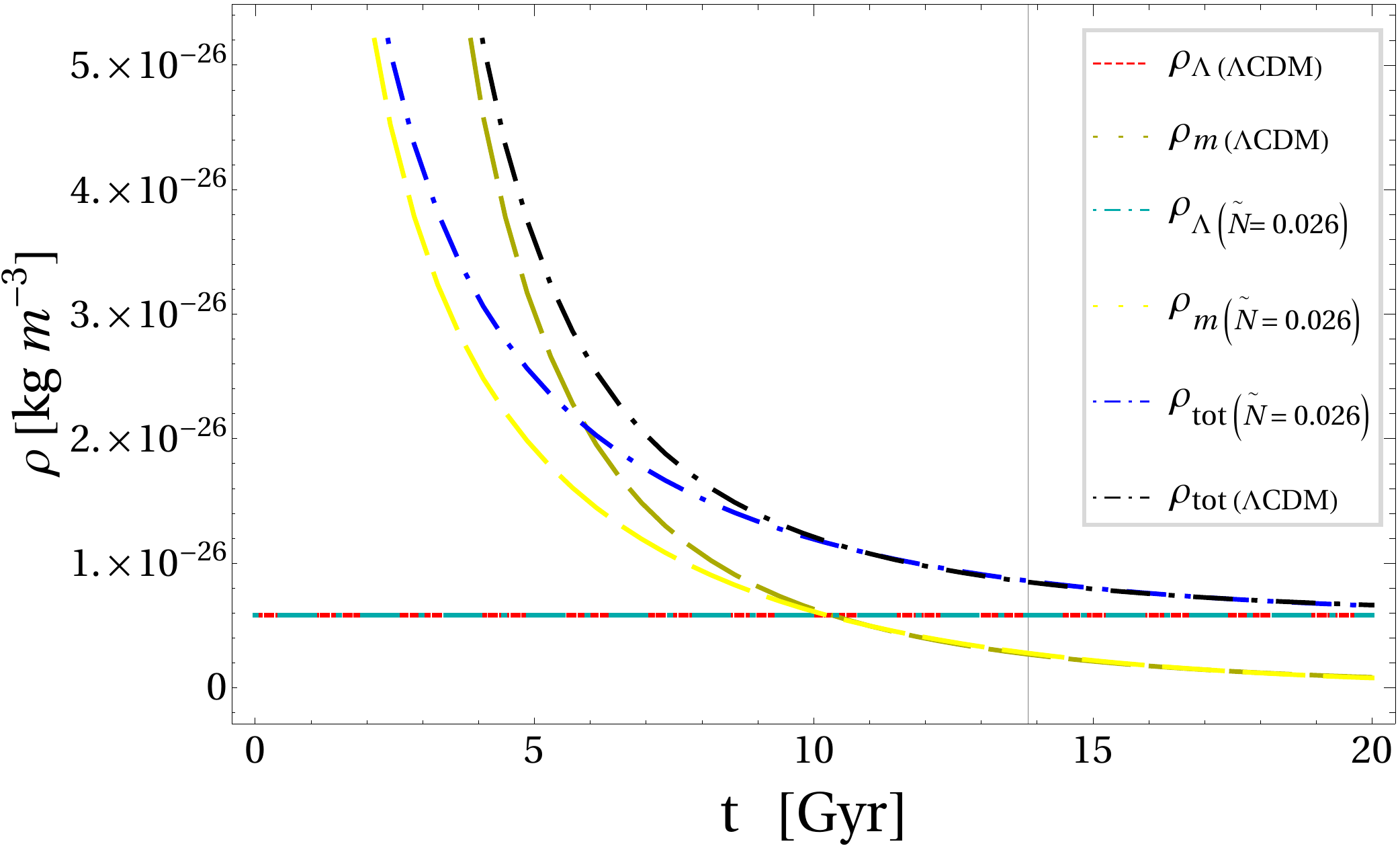}
    \subcaption{The time evolution of 
the matter and the dark energy densities of $\Lambda$CDM are
plotted in the green dashed curve and the red dashed line, respectively. The total $\Lambda$CDM's density is plotted by the black dashed curve. The matter and the dark energy densities of the brane are
plotted in the yellow dashed curve and the cyan dashed line, respectively. The total brane's density parameter is plotted by the blue dashed curve.  }
    \label{ED}
  \end{subfigure}
\caption{}     
\end{figure}

In figure (\ref{allDpar1}) we can see that 
the energy density parameter of the brane- universe evolve with time slightly larger than the energy density parameter of $\Lambda$CDM model, but they have the same value at ($t_0$). Also 
the matter density parameter of the brane- universe is larger than the matter density parameter of $\Lambda$CDM model, but they have a similar behavior, where $\Omega_{m (\tilde{N}=0.026)}$ plotted by the yellow dashed curve is dominant during the early times of the universe then around the time beings it decreases less than $\Omega_{\Lambda (\tilde{N}=0.026)}$ till its current value  $\Omega_{m (\Lambda CDM)}$. Finally 
the total density parameter of the brane- universe has a time dependence during the brane's time evolution, but at the universe's current era ($t_0$) it meets the total density parameter of $\Lambda$CDM in $1$. 
%--------------------------------------------------
\vspace{1cm}

Figure (\ref{ED}) shows that the energy density of the dark energy of the brane in the cyan dashed line coincides
on the energy density of the dark energy of $\Lambda$CDM ($\rho_{\Lambda (\Lambda CDM)} \sim 5.8 \times 10^{-27}$ [kg$~\text{m}^{-3}$]).
The brane's matter energy density in the yellow dashed line
and the $\Lambda$CDM's matter energy density in the green dashed curve are dominant during the early times of the universe, 
then they are suppressed less than $\rho_\Lambda $ near the universe's current era. Also the total energy densities of the brane and 
$\Lambda$CDM are shown, where the first is slightly less than the later during the universe's early times, then they meet around 
$t_0$.

\section{The extra dimensions}

The energy scales of the extra dimensions in the theory are in two distinct classes.
There are the 6-dimensions that yield from compactifying $\D=11$ supergravity to $\N=2$ $\D=5$ supergravity over Calabi-Yau threefold. These dimensions can be treated according to the (ADD) scenario, where the size of extra dimensions is related to the Planck mass of our four-dimensional universe $M_{pl}$ and the Planck mass of the extra dimensions $M$ by 
$R \sim M^{-1} \left( \frac{M_{pl}}{M}\right)^\frac{2}{d} \sim 10^\frac{32}{d}. 10^{-17} ~\text{cm}$. To interpret the hierarchy between $M_{pl}$ and the electroweak scale  $M_{EW}$, the fundamental gravity scale of the extra dimension $M$ is pushed to be of order  $M_{EW}$. In this case, the hierarchy problem can be entirely solved by large size extra dimensions.  For $d=6$, $ R \sim  10^{-12} ~\text{cm}$, which is still larger than the electroweak scale $(1 ~\text{TeV}^{-1}\sim 10^{-17} ~ \text{cm})$. 
Gravity is well-tested to distances around $30 ~ \mu\text{m}$, 
where through these ranges it obeys Newton's inverse-square law \cite{Hoyle:2004cw}. Shorter than sub-millimeters deviations from Newtonian gravity may be probed by future experiments. 
The other energy scale in the theory is the scale of the fifth extra dimension where the hypermultiplets and the complex structure moduli of the CY manifold exist. 
The size of the bulk whether being compactified like ADD model or infinitely large like in Randall–Sundrum \RNum{2}( RS \RNum{2}) braneworld model \cite{Randall:1999}. Next we will show how that can be deduced in our model. According to Friedmann equations (\ref{DPE}) we find:

\be 
a(1) ~b(1) = 0.04,
\label{r1}
\ee
\be
a(t_0) ~b(t_0) = 0.09.
\label{rt}
\ee

\begin{figure}[t]
  \begin{subfigure}[t]{.5\linewidth}
    \centering
    \includegraphics[width=1\columnwidth]{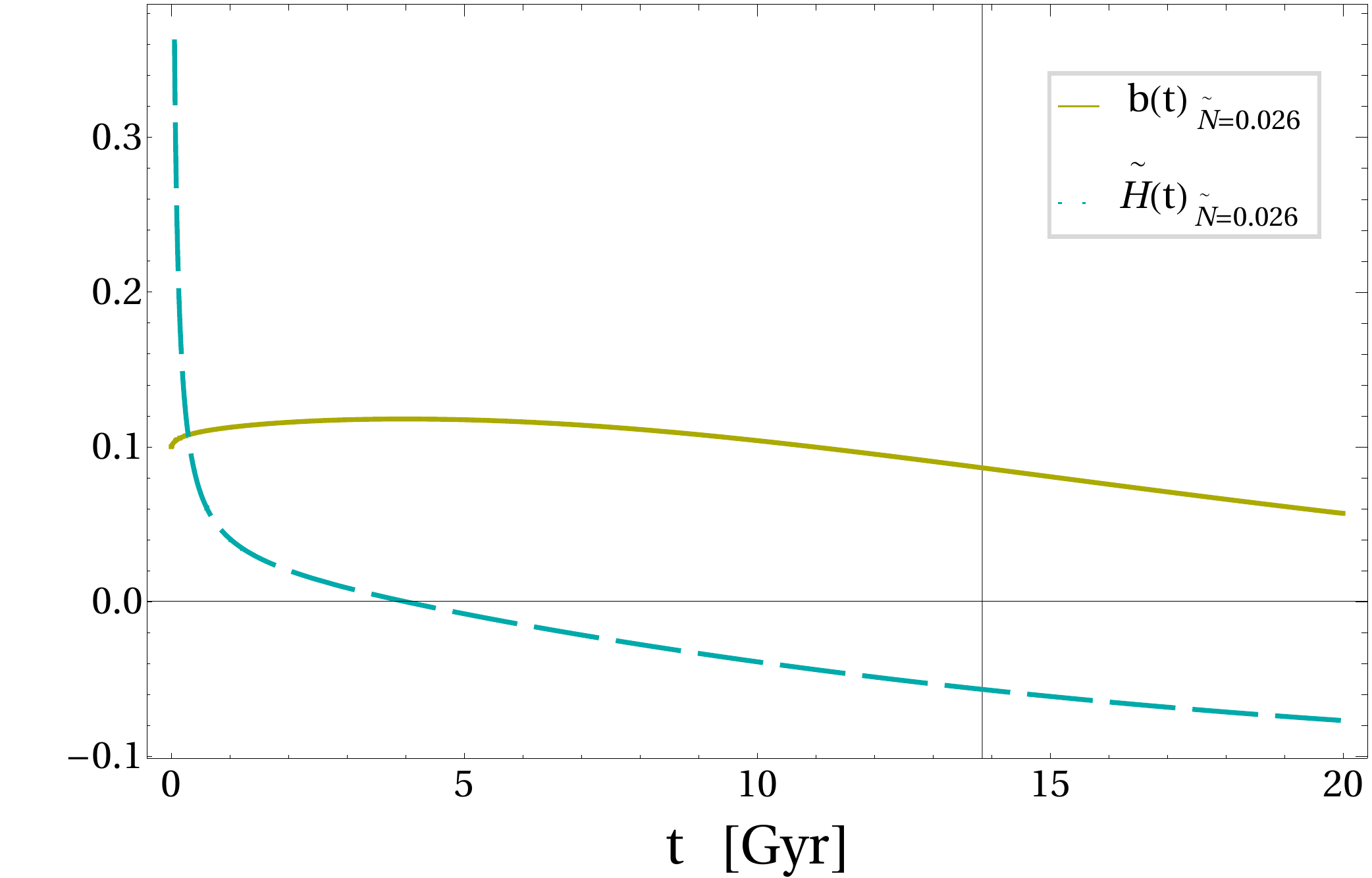}
    \caption{The time evolution of the scale factor and the Hubble parameter of the bulk are represented by the green curve, and the cyan curve, respectively. }
    \label{bb02}
  \end{subfigure} 
\qquad 
  \begin{subfigure}[t]{.5\linewidth}
    \centering
    \includegraphics[width=1\columnwidth]{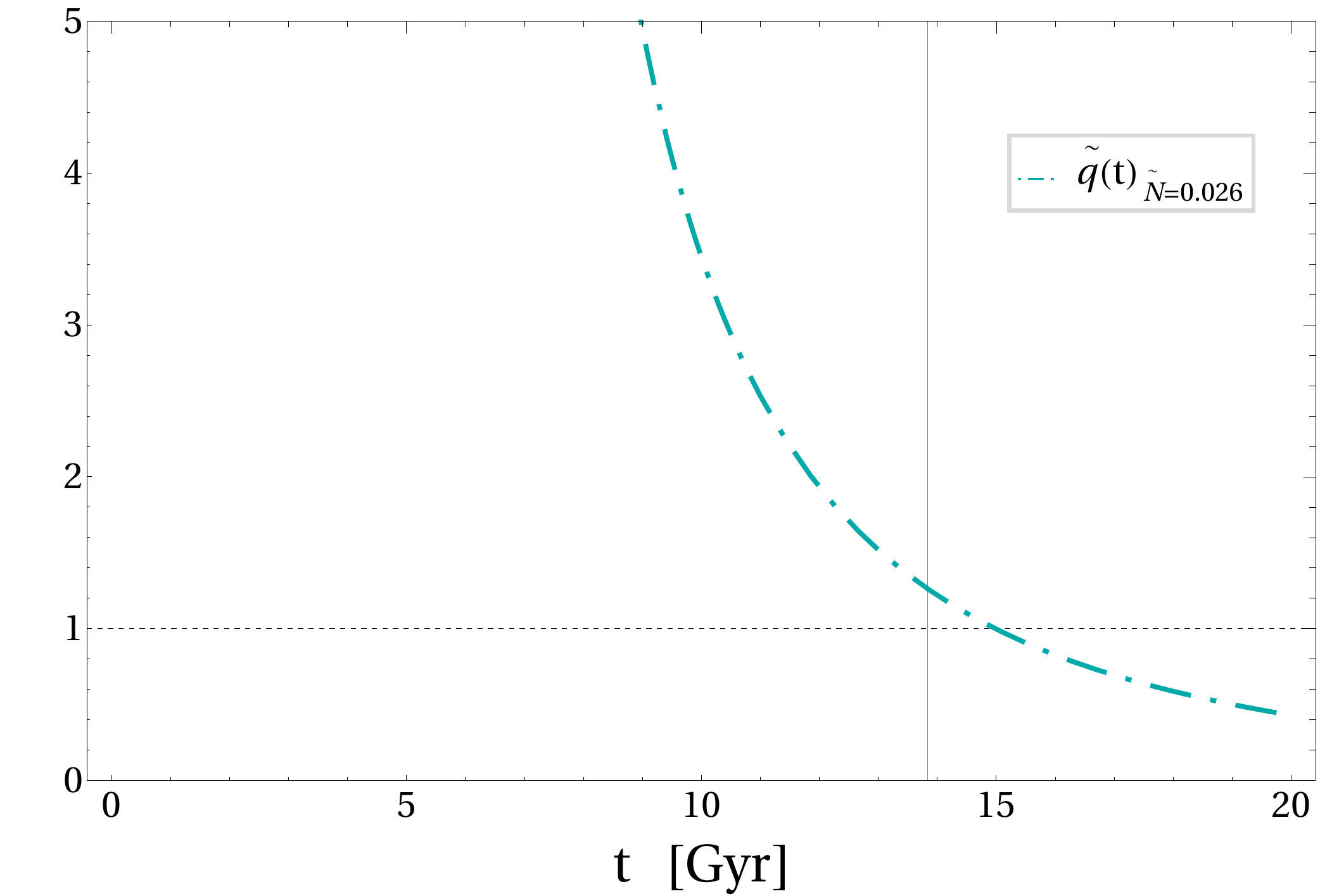}
    \caption{ The deceleration parameter of the bulk is plotted versus time. }
    \label{dec02}
  \end{subfigure}
\caption{}     
\end{figure}

So that $ a(t_0) ~b(t_0) = 2.25 ~ a(1) ~b(1) $. Where $a(1)$ and $b(1)$ are the scale factors of the brane-universe and the bulk, respectively at time $= 1 $ Gyr, and $a(t_0)$ and $b(t_0) $ are the scale factors of the brane-universe and the bulk at the current time of the universe $t_0$. $a(t)$ and $b(t)$ in the Friedmann equations (\ref{DPE}) are the normalized scale factors that relate to the non-normalized scale factors by
$a(t) = \frac{A (t) }{A(t_0)}$ and $b(t) = \frac{B (t) }{B(t_0)}$. Where $ A (t) $ and $ B (t) $ are the non-normalized scale factors of the brane universe and the bulk, respectively in dimensions of meters. $ A (t_0)$ and $ B (t_0)$ are the non-normalized scale factors of the brane-universe and the bulk in dimensions of meters at time  $t_0$. Apply Equ. (\ref{r1}) and (\ref{rt}) on the non-normalized scale factors yields: 
\be
A(t_0) ~B(t_0) = 2.25 ~ A(1) ~B(1) 
\label{rnsc}
\ee

The present observed scale of our universe $ A (t_0)$ is around $10^{26}~ \text{m}$, and assume $ A (1)$ is the initial scale of the brane is around $1 ~ \text{m} \sim 10^{35}$ in Planck units. 
Also, considering the initial scales of the bulk and the brane when forming are equal, then Equ. (\ref{rnsc}) leads to $B(t_0) \sim 
2.25 \times 10^{-26}~ \text{m}$. So the scale of the fifth-dimensional bulk is pretty small but larger than the Planck length $l_p \sim 1.6 \times  10^{-35} ~ \text{m}$, which means physics still holds through the model. Figure (\ref{bb02}) shows the time evolution of the scale factor and the Hubble parameter of the bulk, where $b(t)$ decreases with time, and $\tilde{H}(t)$ rapidly turns to negative values till $t_0$ it's around $\tilde{H}_0 \sim -0.06~ \text {Gyr}^{-2}$. That means the bulk undergoes a decelerated reduction in size. That's also clear in figure (\ref{dec02}) where the bulk’s deceleration parameter 
$ \tilde{q} = -\frac{\ddot{b } b}{\dot{b}^2} $ is plotted versus time. $ \tilde{q}(t_0)$ is ($ > 0$), while the current deceleration parameter of the brane-universe $q(t_0) < 0$. So for the time being when the brane undergoes an accelerated expansion like our universe, the bulk underdogs a decelerated contraction. This bulk's behavior is obvious since all the large vacuum energy density has been transferred to it with a negative sign. 

\section{Conclusion}
The cosmological constant problem can be thought as 
an inquiry of where are the effects of the vacuum energy density on the time-evolution of our universe.
In this work we propose an explanation for that 
by modeling the universe as a 3-brane embedded in the bulk of five-dimensional supergarvity, where the vacuum energy density escapes to the fifth extra-dimension. When solving the modified field equations the variation of the complex structure moduli with respect to the fifth extra-dimension play a crucial role in inducing an effective energy density on the brane equals to the observed value that's responsible for the late-time accelerated expansion of the universe. Also the brane has a cosmic evolution that fits the observations of 
the $\Lambda$CDM model due to the correlation of the brane's scale factor to the moduli flow velocity.  
We find the experimental constrains on the model’s free parameters.
We show that the bulk is an Anti-de Sitter space, its dimension is around  $ (10^{-26}~ \text{m})$ and its negative and large energy density leads it to a decelerated contraction with time. The 6-extra dimensions of $\D=5$ supergravity are copmactified at scale equals $R=(10^{-12}~ \text{m})$, and according to ADD's law within $R$ the scale at which the gravity unifies with the other forces of nature is not $M_{pl}$, but it's $M_{EW}$. So that the weakness of gravity in the brane-universe is due to the existence of unobserved sub-millimeters extra dimensions. The proof of the theory depends on how prices the forthcoming experiments in revealing any modification of gravity from the inverse square law at millimeters scales or less. In conclusion, this can be seen as a complete model that provides a solution for the cosmological constant problem or the cosmological hierarchy and presents an implicit interpretation for the electroweak hierarchy problem. 

%\pagebreak

\section{\it \bf Acknowledgment}

This work is partially supported by the Science, Technology
$\$$ Innovation Funding Authority (STDF) under grant number 48173.

\end{document}